\documentclass[11pt]{article}
\usepackage{epstopdf}
\usepackage{color}
\usepackage{subfigure}
\usepackage{amsmath}
\usepackage{amssymb}
\usepackage{graphicx,color}
\usepackage{cite}
\usepackage{enumerate}
\usepackage{amsthm}
\usepackage{amsfonts,mathrsfs}
\usepackage{geometry}
\usepackage{ifthen}
\usepackage{graphicx}
\usepackage{float}
\usepackage{bm}

\begin{document}

\title{\bf Tighter uncertainty relations based on $(\alpha,\beta,\gamma)$ modified weighted Wigner-Yanase-Dyson skew
information of quantum channels}

\vskip0.1in
\author{\small Cong Xu$^1$, Zhaoqi Wu$^1$\thanks{Corresponding author. E-mail: wuzhaoqi\_conquer@163.com},
Shao-Ming Fei$^{2,3}$\\
{\small\it  1. Department of Mathematics, Nanchang University,
Nanchang 330031, P R China}\\
{\small\it  2. School of Mathematical Sciences, Capital Normal University, Beijing 100048, P R China}\\
{\small\it  3. Max-Planck-Institute for Mathematics in the Sciences,
04103 Leipzig, Germany} }

\date{}
\maketitle

\noindent {\bf Abstract} {\small }\\
We use a novel formation to illustrate the ($\alpha,\beta,\gamma$) modified
weighted Wigner-Yanase-Dyson (($\alpha,\beta,\gamma$) MWWYD) skew
information of quantum channels. By using operator norm
inequalities, we explore the sum uncertainty relations for arbitrary
$N$ quantum channels and for unitary channels. These uncertainty
inequalities are shown to be tighter than the existing ones by a detailed
example. Our results are also applicable to the modified weighted
Wigner-Yanase-Dyson (MWWYD) skew information and the ($\alpha,\gamma$)
modified weighted Wigner-Yanase-Dyson (($\alpha,\gamma$) MWWYD) skew
information of quantum channels as special cases.

\vskip0.2in

\noindent {\bf 1. Introduction}\\\hspace*{\fill}\\
As an extremely important issue in quantum physics, the uncertainty
principle has been widespread concerned since Heisenberg \cite{HW}
proposed the notions of uncertainties in measuring non-commuting
observables. Based on the variance of measurement outcomes the well-known Heisenberg-Robertson
uncertainty relation \cite{RH} says that for
arbitrary two observables $A$ and $B$ with respect to a quantum
state $|\psi\rangle$, one has
\begin{equation}\label{eq1}
\Delta A\Delta B\geq \frac{1}{2}|\langle\psi|[A,B]|\psi\rangle|,
\end{equation}
where $[A,B]=AB-BA$ and $\Delta
M=\sqrt{\langle\psi|{M}^2|\psi\rangle-{\langle\psi|M|\psi\rangle}^2}$
is the standard deviation of an observable $M$.

There are also many ways to describe uncertainty relations, such as entropy
\cite{DD,MHU,WSWA,WSYS,RAE}, variance \cite{GUDDER,DD1,DD2,SL} and
majorization techniques \cite{PZRL,RLPZ,RL,FSGV}. In particular,
the quantum uncertainty can also be characterized by skew
information. The skew information has been initially proposed by
Wigner and Yanase \cite{WY}, termed as {\it Wigner-Yanase}
(WY) skew information. Then a more general quantity has been
suggested by Dyson, called the {\it
Wigner-Yanase-Dyson} (WYD) skew information \cite{WY}. This quantity
has been further generalized in \cite{CL} and termed as {\it
generalized Wigner-Yanase-Dyson} (GWYD) skew information. The
uncertainty relations based on WY skew information, WYD skew
information and GWYD skew information have been studied extensively
\cite{LUO3,YANA1,YANA2,LUO9,WU1,WU2}.

For a quantum state $\rho$ and an observable $A$, Furuichi, Yanagi and  Kuriyama \cite{FURU1}
defined another generalized Wigner-Yanase skew information,
\begin{align}\label{eq2}
\mathrm{K}_{\rho}^{\alpha}(A)=-\frac{1}{2}\mathrm{Tr}
\left(\left[\frac{\rho^\alpha+\rho^{1-\alpha}}{2},A\right]^2\right)
=\frac{1}{2}\left\|\left[\frac{\rho^\alpha+\rho^{1-\alpha}}{2},
A\right]\right\|^{2} ~,\,\,0\leq \alpha \leq 1,
\end{align}
which, called as the {\it weighted Wigner-Yanase-Dyson} (WWYD) skew
information in \cite{WU2}, is different from WYD skew information.
Chen, Liang, Li and Wang \cite{CZL} proposed then a generalized
Wigner-Yanase skew information for arbitrary operator $E$ (not
necessarily Hermitian),
\begin{align}\label{eq3}
\mathrm{K}_{\rho}^{\alpha}(E)=-\frac{1}{2}\mathrm{Tr}\left(\left[\frac{\rho^\alpha
+\rho^{1-\alpha}}{2},E^{\dag}\right]\left[\frac{\rho^\alpha+\rho^{1-\alpha}}{2},E\right]\right)
=\frac{1}{2}\left\|\left[\frac{\rho^\alpha+\rho^{1-\alpha}}{2},
E\right]\right\|^{2} ~,\,\,0\leq \alpha \leq 1,
\end{align}
which is termed as the {\it modified weighted Wigner-Yanase-Dyson}
(MWWYD) skew information in \cite{WU2}.
By replacing the arithmetic mean of $\rho^\alpha$ and
$\rho^{1-\alpha}$ with their convex combination, the {\it
two-parameter extension of the Wigner-Yanase skew information} is
introduced in \cite{Zhang},
\begin{align}\label{eq4}
\mathrm{K}_{\rho,\gamma}^{\alpha}(A)\notag
=&-\frac{1}{2}\mathrm{Tr}\left([(1-\gamma)\rho^\alpha+\gamma\rho^{1-\alpha},A]^{2}\right)\\
=&\frac{1}{2}\left\|\left[(1-\gamma)\rho^\alpha+\gamma\rho^{1-\alpha},
A\right]\right\|^{2} ~,~\,\,0\leq \alpha \leq 1~,\,\,0\leq \gamma
\leq 1,
\end{align}
which is called the {\it $(\alpha,\gamma)$ weighted
Wigner-Yanase-Dyson} ($(\alpha,\gamma)$ WWYD) skew information in
\cite{XWF}. Note that Eq. (\ref{eq4}) reduces to Eq. (\ref{eq2})
when $\gamma=\frac{1}{2}$.

We defined the {\it $(\alpha,\beta,\gamma)$ weighted
Wigner-Yanase-Dyson} ($(\alpha,\beta,\gamma)$ WWYD) skew information
as \cite{XWF},
\begin{align}\label{eq5}
\mathrm{K}_{\rho,\gamma}^{\alpha,\beta}(A)\notag
=&-\frac{1}{2}\mathrm{Tr}([(1-\gamma)\rho^{\alpha}+\gamma\rho^{\beta},A]^{2}
\rho^{1-\alpha-\beta})\\
=&\frac{1}{2}\left\|\left[(1-\gamma)\rho^\alpha+\gamma\rho^\beta,
A\right]\rho^\frac{1-\alpha-\beta}{2}\right\|^{2},~~\alpha,\beta\geq
0,~\alpha+\beta\leq 1,0\leq \gamma \leq 1,
\end{align}
which reduces to Eq. (\ref{eq4}) when $\beta=1-\alpha$.
We also defined the {\it $(\alpha,\beta,\gamma)$ modified
weighted Wigner-Yanase-Dyson} ($(\alpha,\beta,\gamma)$ MWWYD) skew
information with respect to a quantum state $\rho$ and an arbitrary
operator $E$ (not necessarily Hermitian) in \cite{XWF} as
\begin{align}\label{eq6}
\mathrm{K}_{\rho,\gamma}^{\alpha,\beta}(E)\notag
=&-\frac{1}{2}\mathrm{Tr}([(1-\gamma)\rho^{\alpha}+\gamma\rho^{\beta},E^{\dag}]
[(1-\gamma)\rho^{\alpha}+\gamma\rho^{\beta},E]\rho^{1-\alpha-\beta})\\
=&\frac{1}{2}\left\|\left[(1-\gamma)\rho^\alpha+\gamma\rho^\beta,
E\right]\rho^\frac{1-\alpha-\beta}{2}\right\|^{2},~~~\alpha,\beta\geq
0,~\alpha+\beta\leq 1,0\leq\gamma \leq 1,
\end{align}
which is the non-Hermitian extension of the $(\alpha,\beta,\gamma)$
WWYD skew information. Eq. (\ref{eq6}) reduces to Eq.
(\ref{eq10}) in \cite{WU2} when $\gamma=\frac{1}{2}$. When
$\beta=1-\alpha$, we obtain the {\it $(\alpha,\gamma)$ modified
weighted Wigner-Yanase-Dyson} ($(\alpha,\gamma)$ MWWYD) skew
information,
\begin{align}\label{eq7}
\mathrm{K}_{\rho,\gamma}^{\alpha}(E)\notag
=&-\frac{1}{2}\mathrm{Tr}([(1-\gamma)\rho^{\alpha}+\gamma\rho^{1-\alpha},E^{\dag}]
[(1-\gamma)\rho^{\alpha}+\gamma\rho^{1-\alpha},E])\\
=&\frac{1}{2}\left\|\left[(1-\gamma)\rho^\alpha+\gamma\rho^{1-\alpha},
E\right]\right\|^{2},~~\,\,0\leq \alpha \leq 1~,\,\,0\leq \gamma \leq 1,
\end{align}
which is the non-Hermitian extension of the $(\alpha,\gamma)$ WWYD skew information.
It reduces to Eq. (\ref{eq3}) when $\gamma=\frac{1}{2}$.

Quantum channels characterize the general evolutions of quantum
systems \cite{BG,NC}, which play an essential role in quantum
information processing. The uncertainty relations for quantum
channels have been investigated from both the variance-based and
entropic-based uncertainty measure \cite{KMPKR,MASS}. Specifically,
the unitary channels are useful and commonly encountered in
both quantum information theory and quantum computation \cite{NC}.
Uncertainty relations for general unitary channels have been
investigated both theoretically and experimentally
\cite{BS,THSN,BKTN}.
Recently, the sum uncertainty relations for quantum channels have
attracted considerable attention \cite{FSS,ZL,ZWF,CAL,XWF}. Fu, Sun
and Luo \cite{FSS} investigated the uncertainty relations for
two quantum channels based on WY skew information for arbitrary
operators. Afterwards, Zhang, Gao and Yan \cite{ZL} generalized
the uncertainty relations for two quantum channels to arbitrary $N$
quantum channels and proposed tighter lower bounds than the ones
in \cite{FSS} for two quantum channels. Zhang, Wu and Fei \cite{ZWF} proposed new bounds
which are tighter than the results in \cite{ZL}. Cai \cite{CAL}
confirmed that the results in \cite{FSS} also hold for all
metric-adjusted skew information. By employing the norm inequalities
proposed in \cite{ZWF}, we have established sum uncertainty
relations for arbitrary $N$ quantum channels based on
($\alpha,\beta,\gamma$) MWWYD skew information \cite{XWF} .

Following the idea in \cite{SLYS}, the $(\alpha,\beta,\gamma)$ MWWYD
skew information of a state $\rho$ with respect to a channel $\Phi$ has been
defined as \cite{XWF},
\begin{equation}\label{eq8}
\mathrm{K}_{\rho,\gamma}^{\alpha,\beta}(\Phi)=\sum_{i=1}^{n}\mathrm{K}_{\rho,\gamma}^{\alpha,\beta}(E_i),
\end{equation}
where $\alpha,\beta\geq0,~\alpha+\beta\leq 1,0\leq \gamma \leq 1$,
and $E_i(i=1,2,\cdots,n)$ are Kraus operators of the channel $\Phi$,
i.e., $\Phi(\rho)=\sum_{i=1}^{n}E_i\rho E_i^{\dag}$.
Very recently, we provided the following uncertainty relations
for arbitrary $N$ quantum channels $\{\Phi_t\}_{t=1}^N$ with
$\Phi_t(\rho)=\sum_{i=1}^{n}E_{i}^{t}\rho (E_{i}^{t})^\dag,
~t=1,2,\cdots,N$ ($N>2$) \cite{XWF},
\begin{align}\label{eq9}
\sum_{t=1}^{N}\mathrm{K}_{\rho,\gamma}^{\alpha,\beta}(\Phi_t)
\geq& \mathop{\mathrm{max}}\limits_{\pi_t,\pi_s\in S_n}\frac{1}{N-2}\left\{\sum_{1\leq t<s\leq N}\sum_{i=1}^{n}\mathrm{K}_{\rho,\gamma}^{\alpha,\beta}(E_{\pi_{t}(i)}^{t}+E_{\pi_{s}(i)}^{s}) \right.
\nonumber\\
&\left.-\frac{1}{(N-1)^{2}}\left[\sum_{i=1}^{n}\left(\sum_{1\leq t<s\leq N}\sqrt{\mathrm{K}_{\rho,\gamma}^{\alpha,\beta}(E_{\pi_{t}(i)}^{t}
+E_{\pi_{s}(i)}^{s})}\right)^{2}\right]\right\},
\end{align}
\begin{align}\label{eq10}
\sum_{t=1}^{N}\mathrm{K}_{\rho,\gamma}^{\alpha,\beta}(\Phi_t)
\geq& \mathop{\mathrm{max}}\limits_{\pi_t,\pi_s\in S_n}\left\{\frac{1}{N}\sum_{i=1}^{n}\mathrm{K}_{\rho,\gamma}^{\alpha,\beta}
\left(\sum_{t=1}^{N}E_{\pi_{t}(i)}^{t}\right) \right.
\nonumber\\
&\left.+\frac{2}{N^{2}(N-1)}\left[\sum_{i=1}^{n}\left(\sum_{1\leq t<s\leq N}\sqrt{\mathrm{K}_{\rho,\gamma}^{\alpha,\beta}(E_{\pi_{t}(i)}^{t}
-E_{\pi_{s}(i)}^{s})}\right)^{2}\right]\right\},
\end{align}
\begin{align}\label{eq11}
\sum_{t=1}^{N}\mathrm{K}_{\rho,\gamma}^{\alpha,\beta}(\Phi_t)
\geq& \mathop{\mathrm{max}}\limits_{\pi_t,\pi_s\in S_n}\frac{1}{2(N-1)}\left\{\frac{2}{N(N-1)}\left[\sum_{i=1}^{n}\left(\sum_{1\leq t<s\leq N}\sqrt{\mathrm{K}_{\rho,\gamma}^{\alpha,\beta}(E_{\pi_{t}(i)}^{t}
\pm E_{\pi_{s}(i)}^{s})}\right)^2\right] \right.
\nonumber\\
&\left.+\sum_{1\leq t<s\leq N}\sum_{i=1}^{n}\mathrm{K}_{\rho,\gamma}^{\alpha,\beta}(E_{\pi_{t}(i)}^{t}
\mp E_{\pi_{s}(i)}^{s})\right\},
\end{align} 
%
where $\alpha,\beta\geq
0,~\alpha+\beta\leq 1,~0\leq \gamma \leq 1$, $S_n$ is the $n$-element permutation group and $\pi_{t},\pi_{s}\in S_n$ are arbitrary $n$-element permutations.

The remainder of this paper is structured as follows. In Section 2,
we explore the $(\alpha,\beta,\gamma)$ MWWYD skew information-based
sum uncertainty relations for arbitrary $N$ quantum channels.
Especially, we show that when $\beta=1-\alpha$, i.e., when the
$(\alpha,\beta,\gamma)$ MWWYD skew information becomes the
$(\alpha,\gamma)$ MWWYD skew information, our new bounds are tighter
than the existing ones by a detailed example. The uncertainty
relations based on the $(\alpha,\beta,\gamma)$ MWWYD skew
information for unitary channels are discussed in Section 3. We
conclude with a summary in Section 4.
 \\\hspace*{\fill}\\

\noindent {\bf 2. Sum uncertainty relations for arbitrary $N$ quantum channels in
terms of ($\alpha,\beta,\gamma$) MWWYD skew information}
\\\hspace*{\fill}\\
In this section, by using a new formation we explore the uncertainty relations for arbitrary $N$ quantum channels
in terms of the ($\alpha,\beta,\gamma$) MWWYD skew information $\mathrm{K}_{\rho,\gamma}^{\alpha,\beta}(\Phi)$.

Let $\Phi$ be a quantum channel with Kraus representation,
$\Phi(\rho)=\sum_{i=1}^{n}E_i\rho E_i^{\dag}$. Following the idea in
\cite{ZWF}, we define the ($\alpha,\beta,\gamma$) MWWYD skew
information of the channel as,
\begin{align}\label{eq12}
\mathrm{K}_{\rho,\gamma}^{\alpha,\beta}(\Phi)=\frac{1}{2}\mathrm{Tr}(u^\dag u)=\frac{1}{2}\|u\|^2,
\end{align}
where $\alpha,\beta\geq0$, $\alpha+\beta\leq 1$, $0\leq \gamma \leq 1$, $u=(\left[(1-\gamma)\rho^\alpha+\gamma\rho^\beta,
E_1\right]\rho^\frac{1-\alpha-\beta}{2},\left[(1-\gamma)\rho^\alpha+\gamma\rho^\beta \right.$\\ $\left.  ,E_2\right]\rho^\frac{1-\alpha-\beta}{2},\cdots,\left[(1-\gamma)\rho^\alpha+
\gamma\rho^\beta, E_n\right]\rho^\frac{1-\alpha-\beta}{2})$
characterizes some intrinsic features of both the quantum state and
the quantum channel. By employing operator norm inequalities and Eq. (\ref{eq12}), we have the following theorem for arbitrary $N$ quantum channels.
\\\hspace*{\fill}\\
{\bf Theorem 1} Let $\Phi_{1},\cdots,\Phi_N$ be $N$ quantum
channels with Kraus representations
$\Phi_t(\rho)=\sum_{i=1}^{n}E_{i}^{t}\rho (E_{i}^{t})^\dag, ~t=1,2,\cdots,N$ ($N>2$). We have
\begin{align}\label{eq13}
\sum_{t=1}^{N}\mathrm{K}_{\rho,\gamma}^{\alpha,\beta}(\Phi_t)\geq \mathop{\mathrm{max}}\{LB1,LB2,LB3\},
\end{align}
where
\begin{align}\label{eq14}
LB1
&=\mathop{\mathrm{max}}\limits_{\pi_t,\pi_s\in S_n}\frac{1}{N-2}\left\{\sum_{1\leq t<s\leq N}\sum_{i=1}^{n}\mathrm{K}_{\rho,\gamma}^{\alpha,\beta}(E_{\pi_{t}(i)}^{t}+E_{\pi_{s}(i)}^{s}) \right.
\nonumber\\
&\left.-\frac{1}{(N-1)^{2}}\left[\sum_{1\leq t<s\leq N}\sqrt{\sum_{i=1}^{n}\mathrm{K}_{\rho,\gamma}^{\alpha,\beta}(E_{\pi_{t}(i)}^{t}
+E_{\pi_{s}(i)}^{s})}\right]^{2}\right\},
\end{align}
\begin{align}\label{eq15}
LB2
&=\mathop{\mathrm{max}}\limits_{\pi_t,\pi_s\in S_n}\left\{\frac{1}{N}\sum_{i=1}^{n}\mathrm{K}_{\rho,\gamma}^{\alpha,\beta}
\left(\sum_{t=1}^{N}E_{\pi_{t}(i)}^{t}\right) \right.
\nonumber\\
&\left.+\frac{2}{N^2(N-1)}\left[\sum_{1\leq t<s\leq N}\sqrt{\sum_{i=1}^{n}\mathrm{K}_{\rho,\gamma}^{\alpha,\beta}(E_{\pi_{t}(i)}^{t}
-E_{\pi_{s}(i)}^{s})}\right]^{2}\right\},
\end{align}
\begin{align}\label{eq16}
LB3
&=\mathop{\mathrm{max}}\limits_{\pi_t,\pi_s\in S_n}\frac{1}{2(N-1)}\left\{\sum_{1\leq t<s\leq N}\sum_{i=1}^{n}\mathrm{K}_{\rho,\gamma}^{\alpha,\beta}(E_{\pi_{t}(i)}^{t}
\pm E_{\pi_{s}(i)}^{s}) \right.
\nonumber\\
&\left.+\frac{2}{N(N-1)}\left[\sum_{1\leq t<s\leq N}\sqrt{\sum_{i=1}^{n}\mathrm{K}_{\rho,\gamma}^{\alpha,\beta}(E_{\pi_{t}(i)}^{t}
\mp E_{\pi_{s}(i)}^{s})}\right]^2\right\},
\end{align}
$\alpha,\beta\geq 0$, $\alpha+\beta\leq 1$, $0\leq \gamma \leq 1$, $S_n$ is the n-element permutation group and $\pi_{t},\pi_{s}\in S_n$ are arbitrary $n$-element permutations.

\noindent\textit{Proof} The proof is completed directly by using the following inequalities \cite{CB2,ZL,ZWF},
\begin{align*}
\sum_{t=1}^{N} \| u_t\|^2
\geq&\frac{1}{N-2}\left[\sum_{1\leq t<s\leq N} \| u_t+u_s\|^2-\frac{1}{(N-1)^2}
\left(\sum_{1\leq t<s\leq N}\| u_t+u_s\|\right)^2\right],\\
\sum_{t=1}^{N}\|u_t\|^2
\geq&\frac{1}{N}\left\|\sum_{t=1}^{N}u_t\right\|^2
+\frac{2}{N^2(N-1)}\left(\sum_{1\leq t<s\leq
N}\|u_t-u_s\|\right)^2,\\
\sum_{t=1}^{N}\|u_t\|^2
\geq&\frac{1}{2(N-1)}\left[\frac{2}{N(N-1)}\left(\sum_{1\leq t<s\leq N}\| u_t\pm u_s\|\right)^2+\sum_{1\leq t<s\leq N}\|u_t\mp u_s\|^2\right],
\end{align*}
with $\|u_t\|^2=2\mathrm{K}_{\rho,\gamma}^{\alpha,\beta}(\Phi_t)$,
$\|u_t+u_s\|^2=2\sum_{i=1}^{n}
\mathrm{K}_{\rho,\gamma}^{\alpha,\beta}(E_{\pi_{t}(i)}^{t}+E_{\pi_{s}(i)}^{s})$
and $\|u_t-u_s\|^2=2\sum_{i=1}^{n}\mathrm{K}_{\rho,\gamma}^{\alpha,\beta}(E_{\pi_{t}(i)}^{t}-E_{\pi_{s}(i)}^{s})$.
$\Box$

Note that when $\alpha=\beta=\frac{1}{2}$, Theorem 1 reduce to
Theorem 1 in \cite{ZWF}.
As a special case, we use the $(\alpha,\gamma)$ MWWYD skew
information to compare our lower bounds with the existing ones. For
convenience, we denote by $\overline{LB}1$,  $\overline{LB}2$,
$\overline{LB}3$ the right hand sides of (\ref{eq9}), (\ref{eq10})
and (\ref{eq11}), respectively. The following example shows that our results give tighter lower
bounds than $\overline{LB}1$,  $\overline{LB}2$ and $\overline{LB}3$, see Figure 1.
\\\hspace*{\fill}\\

{\bf Example 1} Given a qubit state
$\rho=\frac{1}{2}(\mathbf{1}+\mathbf{r}\cdot\bm{\sigma})$, where $\mathbf{1}$ is the $2\times2$ identity matrix, $\mathbf{r}=(\frac{\sqrt{3}}{2}\cos\theta,\frac{\sqrt{3}}{2}\sin\theta,0)$, $\bm{\sigma}=(\sigma_1,\sigma_2,\sigma_3)$
with $\sigma_j$ $(j=1,2,3)$ the Pauli matrices, and
$\mathbf{r}\cdot\bm{\sigma}=\sum^3_{j=1}r_j\sigma_j$. We consider the following three quantum channels: \\
(i) the amplitude damping channel $\Phi_{AD}$,
\begin{align*}
\Phi_{AD}(\rho)=\sum_{i=1}^2A_i\rho A_i^\dag, \quad
A_1=|0\rangle\langle0|+\sqrt{1-q}|1\rangle\langle1|, \quad A_2=\sqrt{q}|1\rangle\langle1|;
\end{align*}
(ii) the phase damping channel $\Phi_{PD}$,
\begin{align*}
\Phi_{PD}(\rho)=\sum_{i=1}^2B_i\rho B_i^\dag,\quad  B_1=|0\rangle\langle0|+\sqrt{1-q}|1\rangle\langle1|, \quad B_2=\sqrt{q}|0\rangle\langle1|;
\end{align*}
(iii) the bit flip channel $\Phi_{BF}$,
\begin{align*}
\Phi_{BF}(\rho)=\sum_{i=1}^2C_i\rho C_i^\dag,\quad  C_1=\sqrt{q}|0\rangle\langle0|+\sqrt{q}|1\rangle\langle1|, \quad C_2=\sqrt{1-q}(|0\rangle\langle1|+|1\rangle\langle0|)
\end{align*}
with $0\leq q<1$, respectively.

For the case $\alpha=\gamma=\frac{1}{4}$, $q=0.2$ and
$\theta=\frac{\pi}{2}$, we have
$\mathrm{K}_{\rho,\frac{1}{4}}^{\frac{1}{4}}(\Phi_{AD})+\mathrm{K}_{\rho,\frac{1}{4}}^{\frac{1}{4}}
(\Phi_{PD})+\mathrm{K}_{\rho,\frac{1}{4}}^{\frac{1}{4}}(\Phi_{BF})=0.283955$.
The lower bounds $\overline{LB}1$, $\overline{LB}2$ and
$\overline{LB}3$ are 0.275596, 0.2644 and 0.256419, respectively,
and the lower bounds $LB1$, $LB2$ and $LB3$ are 0.260707, 0.26726
and 0.265758, respectively. Obviously, $LB2$ is
tightest among $LB1$, $LB2$ and $LB3$, which is also greater
than $\overline{LB}2$ and $\overline{LB}3$ given in \cite{XWF}.

We also consider the case $\alpha=\gamma=\frac{1}{4}$. For $q=0.4$
and $q=0.9$, the sum and the lower bounds $\overline{LB}1$,
$\overline{LB}2$, $\overline{LB}3$, $LB1$, $LB2$ and $LB3$ are
shown in Figure 1, respectively. Especially, for $q=0.4$,
the sum and the lower bounds are calculated for some special $\theta$,
as listed in Table 1. It can be seen that for $q=0.4$, our
lower bounds $LB2$ and $LB3$ are tighter than $\overline{LB}1$,
$\overline{LB}2$ and $\overline{LB}3$. While for $q=0.9$, our lower
bounds $LB2$ and $LB3$ are tighter than $\overline{LB}1$,
$\overline{LB}2$ and $\overline{LB}3$.
\begin{figure}[H]\centering
\subfigure[]
{\begin{minipage}[Figure-1a]{0.46\linewidth}
\includegraphics[width=1.0\textwidth]{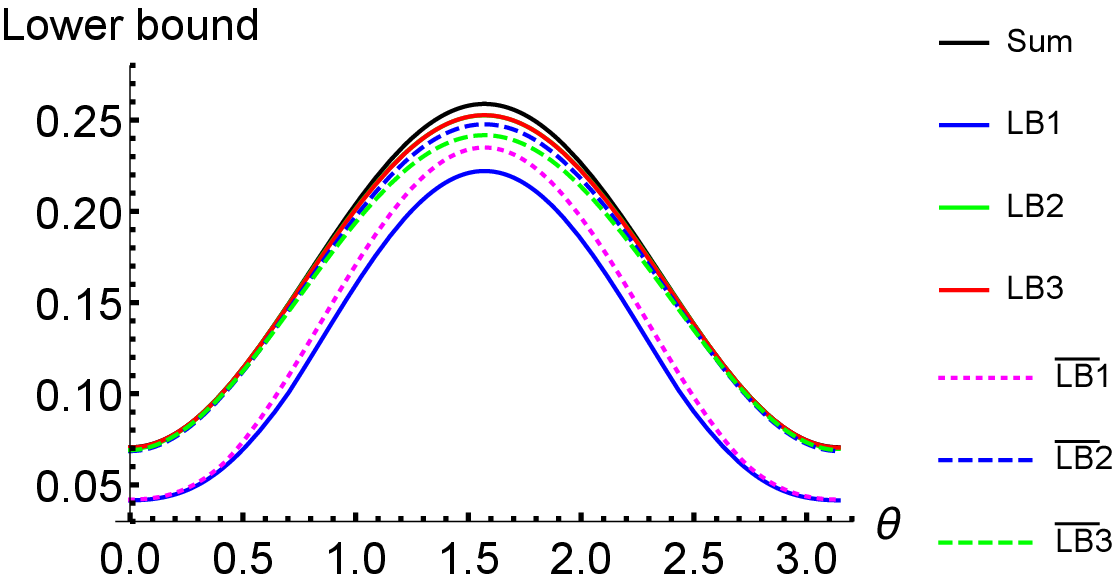}
\end{minipage}}
\subfigure[]
{\begin{minipage}[Figure-1b]{0.46\linewidth}
\includegraphics[width=1.0\textwidth]{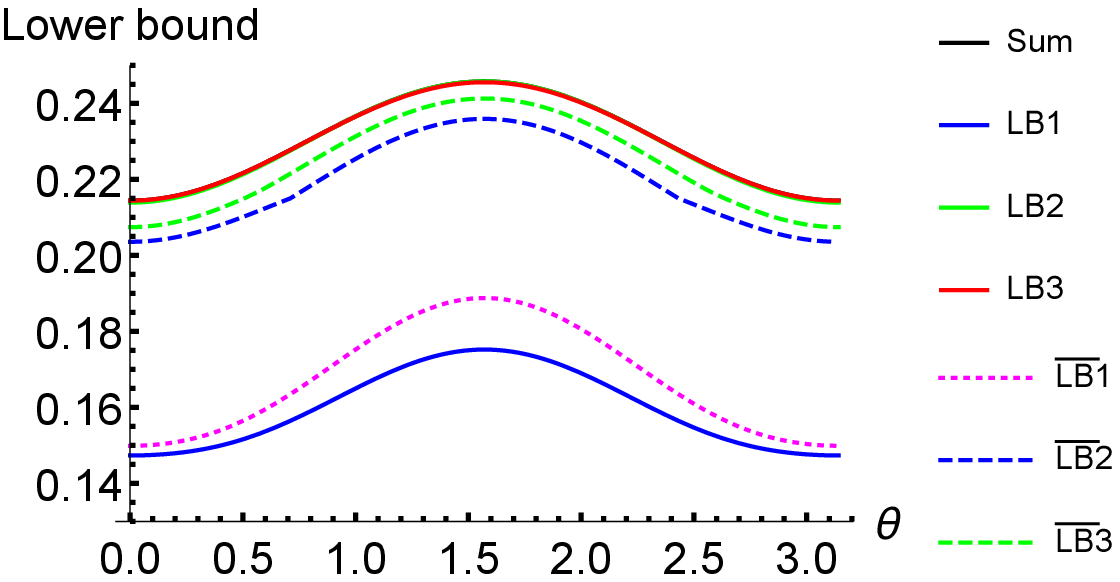}
\end{minipage}}
\caption{{The solid black line represents the sum
$=\mathrm{K}_{\rho,\frac{1}{4}}^{\frac{1}{4}}(\Phi_{AD})
+\mathrm{K}_{\rho,\frac{1}{4}}^{\frac{1}{4}}(\Phi_{PD})+\mathrm{K}_{\rho,\frac{1}{4}}^{\frac{1}{4}}(\Phi_{BF})$.
The solid blue, green and the red lines represent the lower bounds $LB1$, $LB2$ and $LB3$ in Theorem 1,
respectively. The dotted magenta, dashed blue and green lines are for the lower bounds
$\overline{LB}1$, $\overline{LB}2$ and $\overline{LB}3$, respectively. (a) $q=0.4$; (b) $q=0.9$. \label{fig:Fig1}}}
\end{figure}

\begin{center}
Table 1. Comparison among the uncertainty lower bounds
\begin{tabular}{l c c c c c c c r}
\hline
q=0.4 & $\overline{LB}1$ & $\overline{LB}2$ & $\overline{LB}3$ & $LB1$ &$LB2$ & $LB3$& sum \\

$\theta=\pi/2$ & 0.234918  &0.247658  &0.241686 &0.222065 &0.252565 &0.252654  &0.258817  \\

$\theta=\pi/3$ &0.17968 &0.204421 &0.20082  &0.168362  &0.208841 &0.208534 &0.211782 \\

$\theta=\pi/5$ &0.0954994  &0.13303 &0.132687 &0.0879256 & 0.135648 & 0.135459  &0.135679\\

$\theta=\pi/7$ &0.066361   &0.104405 &0.104922  &0.0632504 & 0.106043 & 0.106062 &0.106096 \\
\hline
\end{tabular}
\end{center}

The above results show that Theorem 1 in this paper improve the existing
results ones given in \cite{XWF}.
\\\hspace*{\fill}\\

\noindent {\bf 3. Sum uncertainty relations for $N$ unitary channels
in terms of ($\alpha,\beta,\gamma$) MWWYD skew information}
\\\hspace*{\fill}\\
In this section, we consider sum uncertainty relations for arbitrary
$N$ unitary channels. For a unitary channel $\Phi_U$, we have
$\Phi_U(\rho)=U\rho U^{\dag}$. According to Eq. (\ref{eq6}), the ($\alpha,\beta,\gamma$) MWWYD
skew information of an unitary operator $U$ is given by
\begin{align}\label{17}
\mathrm{K}_{\rho,\gamma}^{\alpha,\beta}(U) \notag
=&-\frac{1}{2}\mathrm{Tr}([(1-\gamma)\rho^{\alpha}+\gamma\rho^{\beta},U^{\dag}][(1-\gamma)\rho^{\alpha}+\gamma\rho^{\beta},U]\rho^{1-\alpha-\beta})\\
=&\frac{1}{2}\left\|\left[(1-\gamma)\rho^\alpha+\gamma\rho^\beta,
U\right]\rho^\frac{1-\alpha-\beta}{2}\right\|^{2},~~~\alpha,\beta\geq 0,~\alpha+\beta\leq 1,0\leq\gamma \leq 1.
\end{align}
The ($\alpha,\beta,\gamma$) MWWYD skew information of a unitary
channel $\Phi_U$ is defined as
$\mathrm{K}_{\rho,\gamma}^{\alpha,\beta}(\Phi_U)=\mathrm{K}_{\rho,\gamma}^{\alpha,\beta}(U)$.
For simplicity, in the following, we use
$\mathrm{K}_{\rho,\gamma}^{\alpha,\beta}(U)$ to denote the quantity
of the unitary channel $\Phi_U$ determined by $U$.
Similar to the proof of Theorem 1, we can prove the following theorem.
\\\hspace*{\fill}\\
{\bf Theorem 2} Let $U_{1},\cdots,U_N$ be arbitrary $N$ unitary
operators. Then we have
\begin{align}\label{18}
\sum_{t=1}^{N}\mathrm{K}_{\rho,\gamma}^{\alpha,\beta}(U_t)\geq \mathop{\mathrm{max}}\{Lb1,Lb2,Lb3\},
\end{align}
where
\begin{align}\label{eq19}
Lb1
&=\frac{1}{N-2}\left\{\sum_{1\leq t<s\leq N}\mathrm{K}_{\rho,\gamma}^{\alpha,\beta}(U_t+U_s) -\frac{1}{(N-1)^{2}}\left[\sum_{1\leq t<s\leq N}\sqrt{\mathrm{K}_{\rho,\gamma}^{\alpha,\beta}(U_t
+U_s)}\right]^{2}\right\},
\end{align}
\begin{align}\label{eq20}
Lb2
&=\frac{1}{N}\mathrm{K}_{\rho,\gamma}^{\alpha,\beta}
\left(\sum_{t=1}^{N}U_t\right)+\frac{2}{N^2(N-1)}\left[\sum_{1\leq t<s\leq N}\sqrt{\mathrm{K}_{\rho,\gamma}^{\alpha,\beta}(U_t
-U_s)}\right]^{2},
\end{align}
\begin{align}\label{eq21}
Lb3
&=\mathop{\mathrm{max}}\limits_{x\in\{0,1\}} \frac{1}{2(N-1)}\left\{\sum_{1\leq t<s\leq N}\mathrm{K}_{\rho,\gamma}^{\alpha,\beta}(U_t
+(-1)^{x}U_s)\right.
\nonumber\\
&\left.+\frac{2}{N(N-1)}\left[\sum_{1\leq t<s\leq N}\sqrt{\mathrm{K}_{\rho,\gamma}^{\alpha,\beta}(U_t
+(-1)^{x+1}U_s)}\right]^2\right\}
\end{align}
and $x\in\{0,1\}$, $\alpha,\beta\geq0,~\alpha+\beta\leq 1,~0\leq \gamma \leq 1$.

Note that (\ref{eq19}), (\ref{eq20}) and (\ref{eq21}) of Theorem 2
reduce to (\ref{eq13}), (\ref{eq14}) and (\ref{eq15}) in \cite{ZWF}
when $\alpha=\gamma=\frac{1}{2}$, respectively.
\\\hspace*{\fill}\\
{\bf Example 2} Given a qubit state $\rho=\frac{1}{2}(\mathbf{1}+\mathbf{r}\cdot\bm{\sigma})$ with
$\mathbf{r}=(\frac{\sqrt{2}}{2}\cos\theta,\frac{\sqrt{2}}{2}\sin\theta,0)$.
Consider the following three unitary operators,
$$
U_1=e^{\frac{i\pi\sigma_1}{8}}=\left(\begin{matrix}
\cos\frac{\pi}{8}\ i\sin\frac{\pi}{8}\\
i\sin\frac{\pi}{8}\ \cos\frac{\pi}{8}
\end{matrix}
\right),
U_2=e^{\frac{i\pi\sigma_2}{8}}=\left(\begin{matrix}
\cos\frac{\pi}{8}\ \sin\frac{\pi}{8}\\
-\sin\frac{\pi}{8}\ \cos\frac{\pi}{8}
\end{matrix}
\right),
U_3=e^{\frac{i\pi\sigma_3}{8}}=\left(\begin{matrix}
e^{i\frac{\pi}{8}} \quad 0\\
\ 0\  -e^{i\frac{\pi}{8}}
\end{matrix}
\right),
$$
which correspond to the rotations around the $z$ axis of the Bloch sphere. When $\beta=1-\alpha$, i.e.,
when the $(\alpha,\beta,\gamma)$ MWWYD skew information reduces to the $(\alpha,\gamma)$ MWWYD skew information,
the comparison among the lower bounds of Theorem 2 is presented in Figure 2, from which one sees that
the lower bound $Lb3$ is tighter than $Lb2$ and $Lb1$ in this case.
\begin{figure}[ht]\centering
{\begin{minipage}[Figure-2]{0.5\linewidth}
\includegraphics[width=0.95\textwidth]{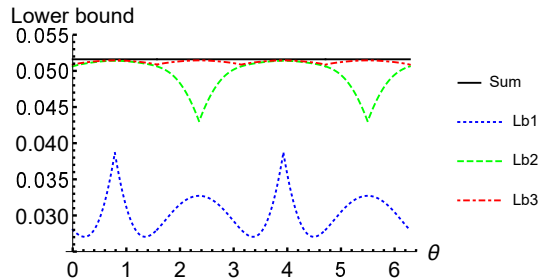}
\end{minipage}}
\caption{For $\alpha=\gamma=\frac{1}{4}$, {the solid black curve represents the sum $=\mathrm{K}_{\rho,\frac{1}{4}}^{\frac{1}{4}}(U_1)+\mathrm{K}_{\rho,\frac{1}{4}}^{\frac{1}{4}}(U_2)+\mathrm{K}_{\rho,\frac{1}{4}}^{\frac{1}{4}}(U_3)$. The dotted blue, dashed green and dot-dashed red curves represent $Lb1$, $Lb2$ and $Lb3$, respectively. \label{fig:Fig3}}}
\end{figure}

\vskip0.1in

\noindent {\bf 4. Conclusions}\\\hspace*{\fill}\\
We have studied the sum uncertainty relations for $N$ quantum channels
based on the $(\alpha,\beta,\gamma)$ MWWYD skew information. By detailed example
it has been shown that our uncertainty inequalities are tighter than the existing ones.
Since the MWWYD skew information and ($\alpha,\gamma$) MWWYD skew information
are two special cases of the $(\alpha,\beta,\gamma)$ MWWYD skew
information, our results are also valid for the MWWYD skew information and the
($\alpha,\gamma$) MWWYD skew information.
Finally, we have also explored sum uncertainty relations for unitary
channels. These results may shed some new light on the
study of skew information-based sum uncertainty relations for quantum channels.
\vskip0.1in

\noindent

\subsubsection*{Acknowledgements}
This work was supported by National Natural Science Foundation of
China (Grant Nos. 12161056, 12075159, 12171044); Jiangxi Provincial
Natural Science Foundation (Grant No. 20202BAB201001); Beijing
Natural Science Foundation (Grant No. Z190005); Academy for
Multidisciplinary Studies, Capital Normal University; Shenzhen
Institute for Quantum Science and Engineering, Southern University
of Science and Technology (Grant No. SIQSE202001); the Academician
Innovation Platform of Hainan Province.

\subsubsection*{Conflict of interest}
\small {The authors declare that they have no conflict of interest.}




\begin{thebibliography}{S2}
\bibitem{HW} Heisenberg W 1927 \"Uber den anschaulichen Inhalt der
quantentheoretischen Kinematik und Mechanik Z. {\it Phys.}
\textbf{43} 172
\bibitem{RH}Robertson H P 1929 The uncertainty principle {\it Phys. Rev.} \textbf{34} 163
\bibitem{DD}Deutsch D 1983 Uncertainty in quantum measurements {\it Phys. Rev. Lett.} \textbf{50} 631
\bibitem{MHU} Maassen H and Uffink J B M 1988 Generalized entropic uncertainty relations {\it Phys. Rev. Lett.} \textbf{60} 1103
\bibitem{WSWA} Wehner S and Winter A 2010 Entropic uncertainty relations-a survey {\it New J. Phys.} \textbf{12} 025009
\bibitem{WSYS} Wu S, Yu S and M${\o}$lmer K 2009 Entropic uncertainty relation for mutually unbiased bases {\it Phys. Rev. A} \textbf{79} 022104
\bibitem{RAE} Rastegin A E 2013 Uncertainty relations for MUBs and SIC-POVMs in terms of generalized entropies {\it Eur. Phys. J. D} \textbf{67} 269

\bibitem{GUDDER} Gudder S 2007 Operator probability theory {\it Int. J. Pure Appl. Math.} \textbf{39} 511

\bibitem{DD1}Dou Y and Du H 2013 Generalizations of the Heisenberg and Schr\"{o}dinger uncertainty relations {\it J. Math. Phys.} \textbf{54} 103508

\bibitem{DD2} Dou Y and Du H 2014 Note on the Wigner-Yanase-Dyson skew information {\it Int. J. Theor. Phys.} \textbf{53} 952
\bibitem{SL} Sun Y and Li N 2021 The uncertainty of quantum channels in terms of variance Quantum Inf. Process. \textbf{20} 25
\bibitem{RLPZ} Rudnicki {\L}, Pucha{\l}a Z and Zyczkowski K 2014 Strong majorization entropic uncertainty relations {\it Phys. Rev. A} \textbf{89} 052115


\bibitem{PZRL} Pucha{\l}a Z, Rudnicki {\L} and Zyczkowski K 2013 Majorization entropic uncertainty relations {\it J. Phys. A: Math. Theor.} \textbf{46} 272002

\bibitem{RL} Rudnicki {\L} 2015 Majorization approach to entropic uncertainty relations for coarse-grained observables {\it Phys. Rev. A} \textbf{91} 032123
\bibitem{FSGV} Friedland S, Gheorghiu V and Gour G 2013 Universal uncertainty relations {\it Phys. Rev. Lett.} \textbf{111} 230401


\bibitem{WY} Wigner E P and Yanase M M 1963 Information contents of distributions {\it Proc. Natl. Acad. Sci.} \textbf{49} 910-918

\bibitem{CL} Chen P and Luo S 2007 Direct approach to quantum extensions of Fisher information {\it Front. Math.} \textbf{2} 359





\bibitem{LUO3} Luo S and Zhang Q 2004 On skew information {\it IEEE Trans. Inf. Theory } \textbf{50} 1778
\bibitem{LUO9} Cai L and Luo S 2008 On convexity of generalized Wigner-Yanase-Dyson information {\it Lett. Math. Phys.} \textbf{83} 253

\bibitem{YANA1} Yanagi K 2010 Uncertainty relation on Wigner-Yanase-Dyson skew information {\it J. Math. Anal. Appl.} \textbf{365} 12

\bibitem{YANA2} Yanagi K 2010 Wigner-Yanase-Dyson skew information and uncertainty relation {\it J. Phys. Conf. Ser.} \textbf{201} 012015

\bibitem{WU1}Wu Z, Zhang L, Fei S-M and Li-Jost X 2020 Coherence and complementarity based on modified generalized skew information {\it Quantum Inf. Process.} \textbf{19} 154
\bibitem{WU2}Wu Z, Zhang L, Wang J, Li-Jost  X, and Fei S-M 2020 Uncertainty relations based on modified Wigner-Yanase-Dyson skew information {\it Int. J. Theor. Phys.} \textbf{59} 704


\bibitem{FURU1} Furuichi S, Yanagi K and Kuriyama K 2009 Trace inequalities on a generalized Wigner-Yanase skew information {\it J. Math. Anal. Appl.} \textbf{356} 179
\bibitem{CZL} Chen Z, Liang L, Li H and Wang W 2016 Two generalized Wigner-Yanase skew information and their uncertainty relations {\it Quantum Inf. Process.} \textbf{15} 5107

\bibitem{Zhang} Zhang Z 2021 Trace inequalities based on two-parameter extended Wigner-Yanase skew information {\it J. Math. Anal. Appl.} \textbf{497} 124851

\bibitem{XWF} Xu C, Wu Z and Fei S-M 2022 Sum uncertainty relations based on $(\alpha,\beta,\gamma)$ weighted Wigner-Yanase-Dyson skew information {\it Int. J. Theor. Phys.} \textbf{61} 185


\bibitem{NC} Nielson M A and Chuang I L 2010
 {\it Quanutm Computation and Quantum Information} (Cambridge: Cambridge University Press) 

\bibitem{BG}Busch P, Grabowski M and Lahti P 1997 {\it Operational Quantum Physics} 2nd edn (Berlin: Springer)
\bibitem{KMPKR} Krishna M and Parthasarathy K R 2002 An entropic uncertainty principle for quantum measurements {\it Sankhy\={a} A} \textbf{64} 842
\bibitem{MASS} Massar S 2008 Uncertainty relations for positive-operator-valued measures {\it Phys. Rev. A} \textbf{76} 042114
\bibitem{BS}Bagchi S and Pati A K 2016 Uncertainty relations for general unitary operators {\it Phys. Rev. A} \textbf{94} 042104
\bibitem{THSN} Tajima H, Shiraishi N and Saito K 2018 Uncertainty relations in implementation of unitary operations {\it Phys. Rev. Lett.} \textbf{121} 110403
\bibitem{BKTN}Bong K-W, Tischler N, Patel R B, Wollmann S, Pryde G J and Hall M J W 2018 Strong unitary and overlap uncertainty relations: theory and experiment {\it Phys. Rev. Lett.} \textbf{120} 230402

\bibitem{FSS} Fu S, Sun Y and Luo S 2019 Skew information-based uncertainty relations for quantum channels {\it Quantum Inf. Process.} \textbf{18} 258

\bibitem{ZL} Zhang L, Gao T and Yan F 2021 Tighter uncertainty relations based on Wigner-Yanase skew information for observables and channels {\it Phys. Lett. A} \textbf{387} 127029

\bibitem{ZWF} Zhang Q, Wu J and Fei S-M 2021 A note on uncertainty relations of arbitrary $N$ quantum channels {\it Laser Phys. Lett.} \textbf{18} 095204

\bibitem{CAL} Cai L 2021 Sum uncertainty relations based on metric-adjusted skew information {\it Quantum Inf. Process.} \textbf{20} 72
\bibitem{SLYS} Luo S and Sun Y 2018 Coherence and complementarity in state-channel interaction {\it Phys. Rev. A.} \textbf{98} 012113
\bibitem{CB2}Chen B and Fei S-M 2015 Sum uncertainty relations for arbitrary $N$ incompatible observables {\it Sci. Rep.} \textbf{5} 14238


\end{thebibliography}
\end{document}